\documentclass{article}
\usepackage[english]{babel}
\usepackage[letterpaper,top=2cm,bottom=2cm,left=2cm,right=2cm,marginparwidth=1.75cm]{geometry}

\usepackage{bbold}
\usepackage[usenames, dvipsnames]{color}
\definecolor{mygray}{gray}{0.6}
\def\bSig\mathbf{\Sigma}

\usepackage[figuresright]{rotating}
\usepackage{amsmath}
\usepackage{amssymb}
\usepackage{authblk}
\usepackage{graphicx}
\usepackage{subcaption}
\usepackage{bm}
\usepackage{float}
\usepackage{array}
\usepackage[section]{placeins}
\usepackage{multirow}
\usepackage{multicol}
\usepackage[round]{natbib}
\usepackage{makecell}
\usepackage[colorlinks=true, linkcolor=blue, citecolor=blue, urlcolor=blue]{hyperref}
\usepackage[usenames, dvipsnames]{color}
\definecolor{mygray}{gray}{0.6}
\def\bSig\mathbf{\Sigma}

\begin{document}

\title{Zero \& $N$-inflated overdispersed binomial models for sum-constrained Poisson count processes}
\author[1]{James Sweeney\thanks{Corresponding author: \texttt{james.a.sweeney@ul.ie}}}
\author[2]{John Haslett}
\author[3]{Dipankar Bandyopadhyay}
\author[4]{Michael Fop}
\author[4]{Andrew Parnell}

\affil[1]{Department of Mathematics and Statistics, University of Limerick, Ireland}
\affil[2]{School of Computer Science and Statistics, Trinity College Dublin, Dublin D2, Ireland}
\affil[3]{Department of Biostatistics, School of Public Health, Virginia Commonwealth University,
USA}
\affil[4]{School of Mathematics and Statistics, UCD, Dublin D4, Ireland}

\date{}
\maketitle

\begin{abstract}

A frequent challenge encountered with compositional ecological data is how to interpret and model data with a high proportion of zeros and $N$'s. Such data frequently occur in ecological applications where counts of species are collected until a pre-specified total imposed (typically) by sampling cost is reached. In the bivariate count (two-species) setting we focus on in this article, zero-inflation of one species will result in $N$-inflation of the other. This can lead to species absence being attributed to an unsuitable habitat as opposed to missingness by chance. Similarly, an excess of $N$'s will lead to misleading inferences about habitat preference and abundance estimates. 
Our contribution is to identify that two independent zero-inflated Poisson processes subject to a sum constraint provide a novel biologically-motivated generating mechanism for the occurrence of binomial count data exhibiting zero and $N$-inflation. We identify an extension to the model to capture additional overdispersion within the data resulting in a novel zero and $N$-inflated beta-binomial model. 
We consider two motivating datasets, one involving a pesticide treatment for an invasive species, and a second involving the abundance of two plant species. We demonstrate that incorporation of covariates in each case enable learning about sources of zero and $N$-inflation as well as abundance. We show that the models result in improved understanding of underlying biological processes as well as improved predictive performance.



\end{abstract}




\section{Introduction}
\label{s:intro}

Count data with many zeros are common in a wide variety of disciplines with sample applications including: modelling defects in manufacturing~\citep{Lambert}, repeated measures studies in biology~\citep{Hall},  the identification of disease risk factors in oral health~\citep{Bandyopadhyay}, and  statistical climatology with a focus on abrupt historical climate change~\citep{Haslett, Parnell}. The excess of zeros in each of these applications has prompted the development of zero-inflated models such as the zero-inflated Poisson (ZIP) distribution of~\citet{Lambert}, the zero-inflated binomial (ZIB) distribution of~\citet{Hall}, and the spatial zero-inflated beta-binomial (ZIBB) distribution of~\citet{Bandyopadhyay}. In applications where there is a clear sequential order to the decision making process of the generated data, hurdle models may be appropriate~\citep{Mullahy1986}. With a hurdle model the values greater than $0$ are modelled using a truncated statistical distribution. As none of the datasets considered in this article align with this framework of sequential decision making processes generating data, we do not discuss this class of models further. More recently, it has been recognised that compositional data exhibiting signs of excess zeros may also exhibit $N$ (or endpoint) inflation, where in addition to an excess of zeros, an excess of $N$'s for elements of the groups comprising the compositional counts is observed \citep{Tian2015}. Both \cite{Royle2006} and \cite{Royle} detail via simulation studies the result of failing to account for an excess of zeros or $N$'s in ecological trials; estimates of species prevalence will be biased downwards in the presence of zero-inflation, or biased upwards in the case of $N$-inflation. 

To emphasize the prevalence of this zero/$N$ problem and the impact of failing to account for it, we present two separate highly-cited studies in which we identify zero and $N$-inflation within the collected data. The first example is the original \cite{Hall} article where the zero-inflated binomial (ZIB) distribution is introduced. We show that this model fits the data of the case study poorly in the published article, and that the zero and $N$-inflated version provides a superior fit and more natural interpretation of the results. The second example concerns an application in statistical climatology using data from \cite{Haslett} where there is interest in utilising fossil pollen data gathered from lake sediment to determine the prevailing climate at a given location at the time of fossil pollen deposition. We illustrate that the use of a zero-inflated model results in misleading inferences on the preferred climate range of a plant genus, and in erroneous inferences on the predicted climate corresponding to fossil pollen leading to vastly differing inferences being drawn on the climate of the past. 

While this article is focused on modelling discrete count applications where the sum constraint is known, a number of authors have considered models for proportion data ($p$ $\in$ $[0,1]$) exhibiting multimodality at the extremes of the distribution. \citet{Ospina2012} propose and study a general class of beta regression models for continuous proportions when the data contain many zeros or ones, in addition to fractional values between these extremes. This work has been shown to have applications in many fields including credit scoring \citep{Louzada2017, Pereira2013}, illness monitoring in terms of flu trends \citep{guolo2014},  as well as having several applications in ecology \citep{Joseph2016, Wright2017}. More recent developments have focused on extending this set of models to financial applications \citep{Tomarchio2019}. 

The focus of this article is on applications consisting of discrete sum constrained bivariate count observations - it must be noted that there is substantially more information available in datasets where raw counts of species are available as opposed to proportion data. This can be intuited from the calculation of classical confidence intervals for proportions based on binomial count data where the sum total is known. \cite{Deng2015} highlight the issue of zero/one-inflation in binary counts giving the example of epidemiological studies, where the incidence of an infective disease in some families is either zero or 100\% during a period of infection. They develop a series of score tests for testing whether endpoint-inflation exists, and apply it to the whitefly dataset of \citet{Hall}. \cite{Tian2015} extend the zero-inflated Bernoulli distribution to a generalised zero \& endpoint-inflated (\textit{ZEIB}) binomial distribution. Six different count generating representations for zero and endpoint inflated random variables are presented via mixture distributions, and the distributional properties of these generalisations are extensively studied. However, the generating mechanisms are theoretical in nature, and are not motivated by the applications presented, which includes the whitefly dataset of \cite{Hall}. \cite{Tian2015} also provide a number of simulation studies to illustrate the bias of model estimates of zero-inflated only models when not accounting for evident $N$-inflation in datasets. In more recent times, exploration of these models have focused on theoretical aspects and model fitting as opposed to detailed application to data. \cite{Dupuy} explores the large-sample properties of the maximum likelihood estimates which arise from inference in the case of zero and $N$-inflated binomial data. \cite{diallo2019estimation} extend this class of models to consider regression settings where covariates are missing at random. Extensions to multinomial count settings have also been recently considered, for example  \cite{Koslovsky2023} and \cite{menezes2025}, however the focus is once more on the development of theory as opposed to a focus on the insights obtained from applications to data.

Our contributions in this article are twofold, both theoretical and applied. From a theoretical viewpoint we propose an additional data generating mechanism for bivariate count data exhibiting zero and $N$-inflation that is motivated in biological theory, as well as identifying an intuitive extension that allows for excess overdispersion in the observed counts. The first generating mechanism explored is as a mixture distribution,
based along the lines of plausible biological count generating processes within the data. An alternative generating mechanism arises from the consideration of bivariate zero-inflated Poisson count outcomes subject to a potentially varying sum constraint, which illustrates that the observance of both zero and $N$ inflated binomial count outcomes in ecological studies is a completely natural phenomenon. From an applied statistical analysis point of view, our contributions include the presentation of two separate studies where there are substantially differing inferences drawn from the fitting of previous models. In particular, we illustrate that the use of a zero-inflated model results in incorrect inferences on the preferred climate range of a plant genus, resulting in potentially misleading inferences on the predicted climate corresponding to fossil pollen, as used in palaeoclimate reconstruction. This is of great concern given the calibration of future model predictions by applying developed models to historical data for which climate can be inferred.


We structure the article as follows: in Section~\ref{Motivating_Examples} we introduce the whitefly and pollen datasets which motivate our work. In Section~\ref{Natural} we describe how zero and $N$-inflated binomial count data can naturally arise in the mixture model setting, or as a set of sum constrained zero-inflated Poisson counts. We also provide an illustrative example of the erroneous inferences obtained by standard models when applied to these data. In Section~\ref{ZNIB_APP} we present an extension of the ZIB model of \cite{Hall} to consider $N$-inflated counts, as well as introducing a beta-binomial adaptation which is flexible enough to consider additional heterogeneity in collected data. 
In Section~\ref{Applications} we illustrate the superiority of the new class of models with application to the whitefly and pollen examples, and a brief summary is included in Section~\ref{Discussion}. 



\subsection{Motivating Examples}\label{Motivating_Examples}


The whitefly dataset analysed in the ZIB regression paper of \cite{Hall} concerns the application of the insecticide imidacloprid to suppress an invasive pest (whitefly) which affects the growth of poinsettia plants. Six variations of treatments are considered - the application of the insecticide via subirrigation following 0, 1, 2, and 4 days respectively without water, a hand watered treatment, and a control treatment in which no pesticide is applied. The treatments are applied in a randomised complete block design with repeated measures over 12 consecutive weeks. The experimental unit is a trio of poinsettia plants, and 18 such units (54 plants) are randomised to the six treatments in three complete blocks. Clip on cages are attached to each plant, and the response variable we consider concerns the number of whitefly surviving after two days of exposure to each treatment, with 640 observations available in total. The dataset is heavily zero-inflated (53\% of observations are zeros), as well as $N$-inflated (12\% of observations are $N$'s, mostly 10, representing full survival). Hall does not identify this $N$-inflation, only intuiting that the zero-inflation can be separated into two distinct settings, (1) the pesticide is fully effective, resulting is complete elimination of the pest (0 survival), or (2) the pesticide is partially effective, resulting in the death of \textit{some} of the pests. There is no pesticide applied in the control treatment so inflation at $N$ (full survival) is presumably completely natural. However, for the control set there are several zero survival cases (full mortality at treatment end) as well as many cases of non-survival.

\begin{figure}
\centerline{\includegraphics[width=6.5in]{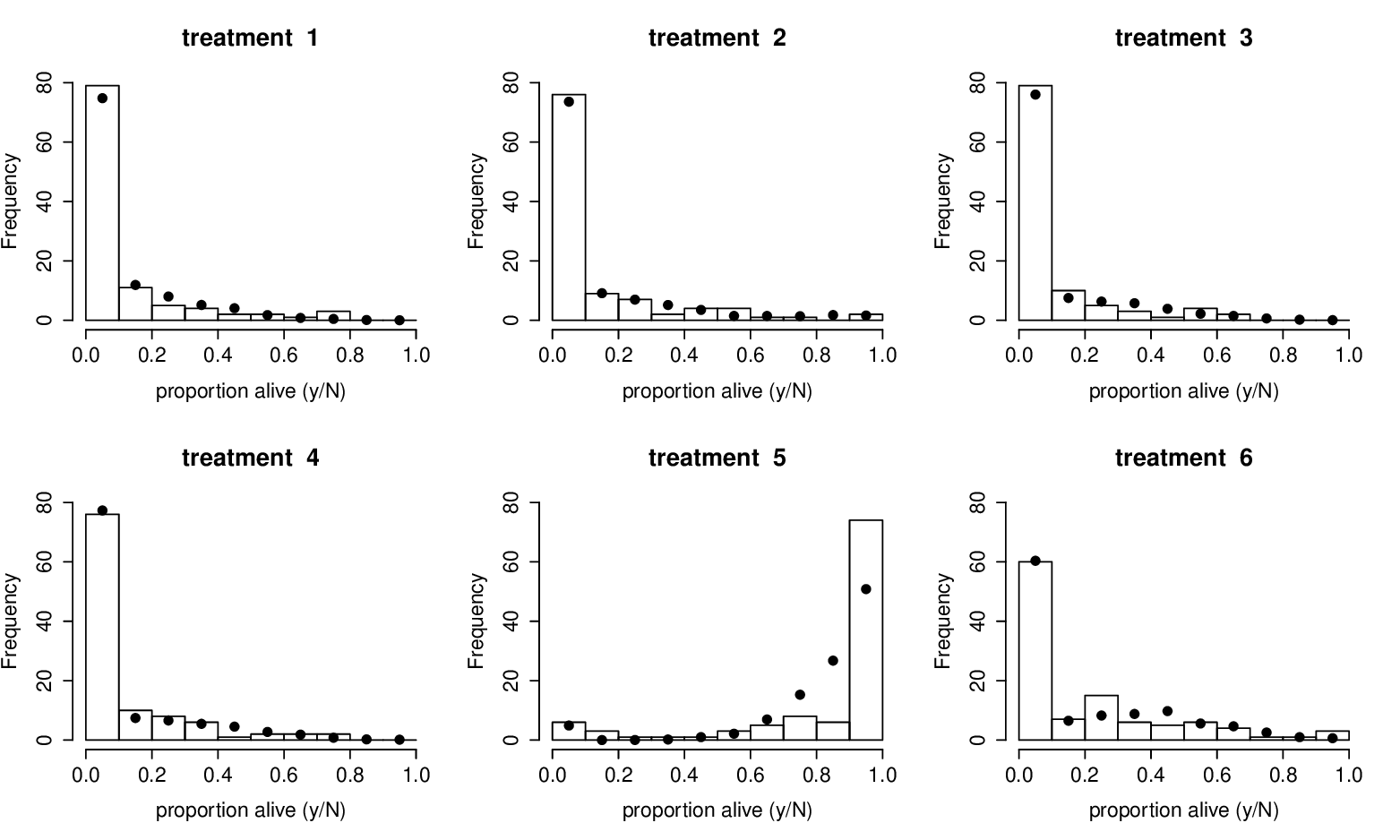}}
\caption{Frequency histogram of observed proportions ($y_i/N_i$) for the live insects with the predicted frequencies ($\bullet$) of the ZIB model of Hall overlain.}
\label{f:Hall}
\end{figure}

In Figure~\ref{f:Hall} we present histograms of the proportion of surviving insects within each treatment category, as well as the predicted survivor proportions of the ZIB model fitted to the survivor outcomes data in \cite{Hall}. Note, treatment 5 concurrently displays signs of both zero and $N$-inflation, i.e. a larger proportion of the whitefly survive than would be expected given the model, with the ZIB model poorly matching the surviving proportions. The poor predictive performance of the \cite{Hall} model is further observed in Table~\ref{Table1}, showing that the number of experiments where the proportion of surviving whitefly is between 80\%-90\% is grossly overestimated, as well as underestimating the number of experiments with full survival. We argue that the \cite{Hall} model is not sufficiently sophisticated to allow for complete \textit{ineffectiveness}, which is observed in the control treatment due to the lack of a pesticide application. We explore this further in Section~\ref{HallApp}, and detail how the strong plant-specific effect identified by \cite{Hall} vanishes once allowances are made for $N$-inflation, and results in changes to the previous study conclusions.

\begin{figure} \centerline{\includegraphics[width=7.5in]{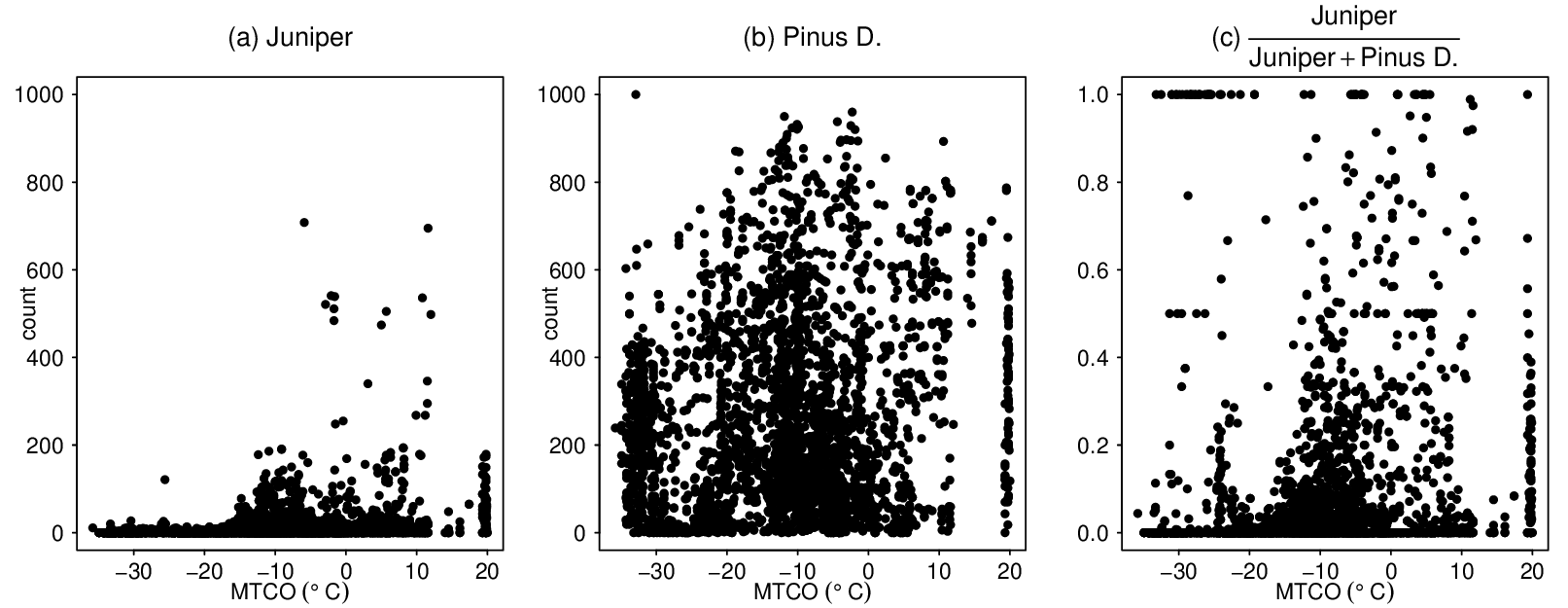}}
\caption{Observed pollen counts for (a) Juniper and  (b) Pinus D. In (c) we plot the proportion of Juniper pollen in terms of the total Juniper and Pinus D. pollen observed at each site.}
\label{f:Pollen}
\end{figure}

The second example concerns a pollen dataset from \citet{Huntley}, analysed further in \cite{Haslett}, and \cite{Parnell}, where the problem of interest is the development of a model linking the pollen composition at a number of sites to a local climate variable. Pollen is an important source of proxy information on climate as pollen is almost ubiquitous, being so geographically widespread, in comparison with alternatives such as isotope information provided by ice sheets for example. In this example we consider the pollen of two similar genera, \textit{Pinus Diploxylon} (henceforth Pinus D.) and that of the \textit{Juniperus} (henceforth Juniper) genus, consisting of pollen counts at $4619$ sites, all in North America. The climate variable of interest is $MTCO$ (the mean temperature of the coldest month), which is a measure for the degree of coldness in winter.  The pollen counts for each genus are negatively correlated as a pre-specified $N_i$ number of pollen spores are counted at each site $i$. There is substantial variation in the $N_i$'s which range in value from 1 to 1000 depending on the prevalence of the pollen of both species at a site. In Figure~\ref{f:Pollen} we observe that the Pinus D. counts at each site typically dominate those of Juniper, with the majority of the Juniper counts zero for values of $MTCO$ less than $-10^{\circ}C$. The Juniper counts exhibit signs of zero-inflation, with $\approx$ 60\% of the Juniper counts being zero. Conversely, in spite of Pinus D. being the dominant genera, approximately 1.5\% of the Pinus D. counts are zero. The data are compositional in nature - the Juniper counts appear to suffer from both zero and $N$-inflation; this is clearly seen in the proportion of Juniper counts observed at each site in Figure~\ref{f:Pollen}. These $N$'s arise as a result of zeros in the Pinus D. counts, with the observed pollen at the site being Juniper only, even at sites where the climate does not appear to be favourable to Juniper.


In Section~\ref{PollenApp} we show that zero-inflated beta-binomial (ZIBB) models for the Juniper counts are unable to account for the additional source of variance provided by the excess of $N$'s, which are due to the absence of Pinus D. pollen at these sites. This results in misleading and non-credible inferences on the pollen-climate relationship for these genera. The fitting of zero-inflated models will result in an overestimation of the response to climate of the Juniper species at low temperatures - the $N$'s observed are a function of the small number of zeros observed for the Pinus D. counts, as opposed to the individual site climate being favourable to Juniper. 

The fitting of models linking present day pollen composition to local climate variables has a further important application in making inferences on past climate, which can be used to assess climate change. Fossil pollen for a variety of plant species, obtained from cores of lake sediment and radiocarbon dated, provides a noisy source of information on $MTCO$ at that location over the past several thousand years. We can use the models fitted to modern pollen data to make quantitative inferences on the $MTCO$ at the time of pollen deposition for these data. \cite{Haslett} provide a detailed account of the data collection and modelling procedures involved, and \cite{Parnell} illustrates how reconstructions of past climate are obtained when fully accounting for temporal uncertainty. Here, we use a left-out portion of the modern pollen counts as a pseudo fossil pollen validation set to assess the predictive performance of the zero-inflated models in comparison to zero and $N$-inflated ones, detailing that the use of zero-inflated models results in implausible statements on past climate.

\section{Biological background and plausibility of zero \& $N$-inflation}\label{Natural}


When ecological data such as counts of species are collected, it is typical to count up to an upper pre-specified bound due to constraints on resources. Here we restrict ourselves to the two species setting, but the arguments generalise to $n >2$.  In the following we consider two generating mechanisms for zero and $N$-inflated binomial count outcomes and illustrate the bias of standard binomial and ZIB models when $N$-inflation is ignored.

First, consider an experiment to collect information on two species A and B, with the counts $X_A$ and $X_B$ constrained to a sum total $N$, i.e. $X_A + X_B = N$. Given $N$, the counts of each species can be considered as a binomial response - given knowledge of $X_A$ and $N$, $X_B$ is explicitly known. However, if we consider a count of zero for each species as having two interpretations, as per \cite{Martins}:

\begin{enumerate}
\item True zero: species does not occur at a site because the habitat is unsuitable or the species has not saturated its entire suitable habitat by chance.
\item False zero: species occurs at a site but is not present during the survey period, or the species occurs and is present but the observer fails to detect it.
\end{enumerate}

The counts for species A may be zero-inflated (due to observer error), or $N$-inflated, equally due to observer error in detecting species B. The reasons (biological or otherwise) for $N$-inflation in one variable are exactly the same as for zero-inflation in the other.  As an illustrative example,we simulate the counts of each species $X_A$ and $X_B$ as independently arising from two separate zero-inflated Poisson distributions \citep{Lambert}, i.e. $x_A \sim \mathrm{ZIP}(\lambda_{X_A},q_{X_A})$ and $ x_B \sim \mathrm{ZIP}(\lambda_{X_B},q_{X_B})$ such that:\\ 
\begin{minipage}{0.45\textwidth}

$$ X_A \sim \left\{ \begin{array}{lcc} 0 & \mbox{with probability} &  q_{X_A} \\ 
\mathrm{Poisson}(\lambda_{X_A}) & \mbox{with probability} & (1-q_{X_A})  
 \end{array} \right.$$
\end{minipage}%
\hfill
\begin{minipage}{0.45\textwidth}
$$ X_B \sim \left\{ \begin {array}{lcc} 0 & \mbox{w. p.} &  q_{X_B} \\ 
\mathrm{Poisson}(\lambda_{X_B}) & \mbox{w. p.} & (1-q_{X_B})  
 \end{array} \right.$$ 
 \end{minipage}\\

\begin{figure}
 \centerline{\includegraphics[scale=.7]{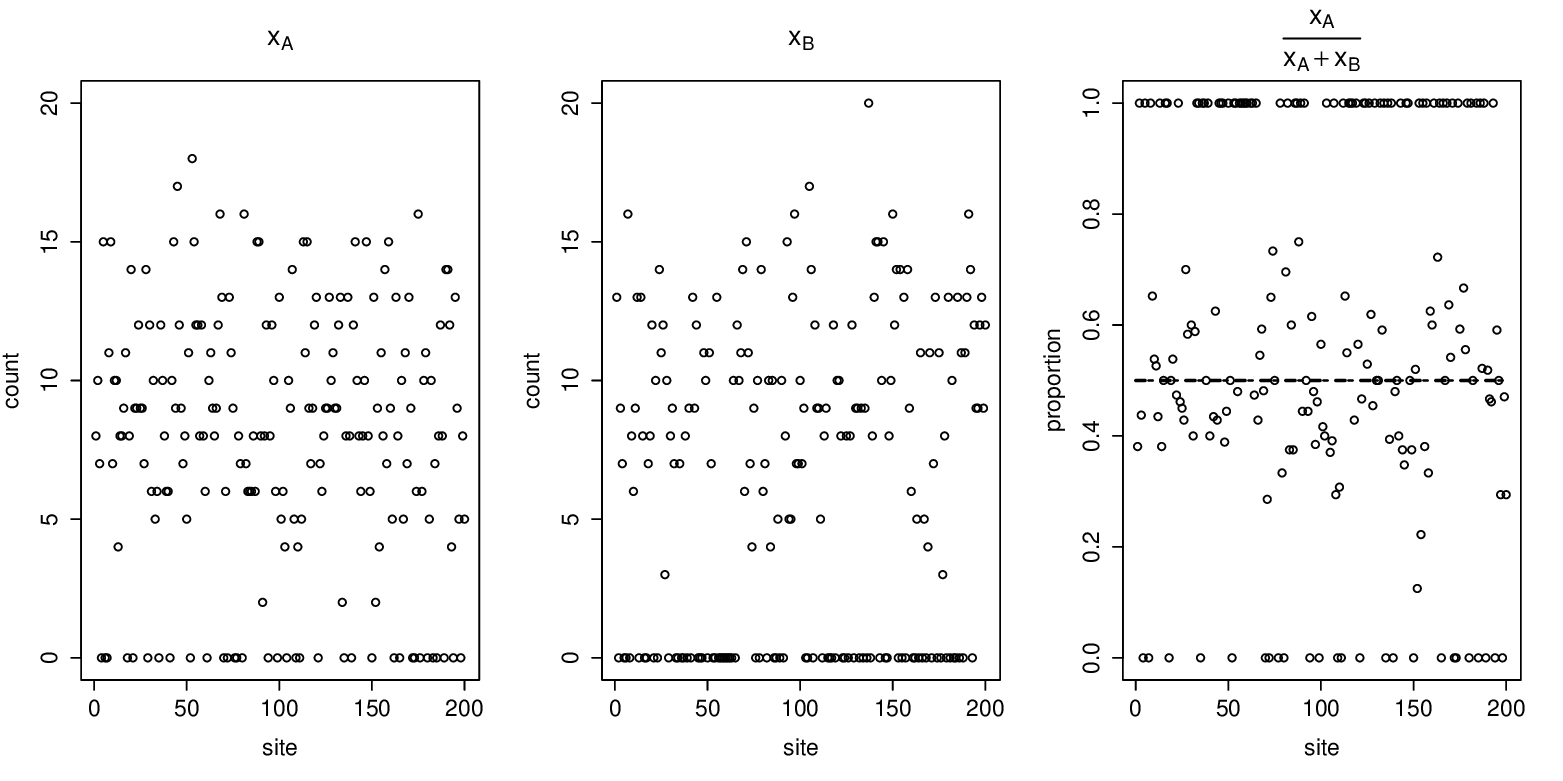}}
\caption{Simulated counts, $x_A$ \& $x_B$, for two independent zero-inflated Poisson count processes. Conditional on their sum total ($x_A$+$x_B$), the proportion of $x_A$ counts is both zero and one inflated. Dashed line at $\frac{\lambda_{X_A}}{\lambda_{X_A}+\lambda_{X_B}}=0.5$ represents the true proportion.}
\label{f:CountSims}
\end{figure}

Suppose that observers at individual sites $i = 1, \ldots, n$ collect data until $N_i$ samples have been collected, where the sum totals $N_i$ vary from site to site. If the counts of species $A$ and $B$ are both zero-inflated, then the sum constrained counts of an individual species will be both zero \& $N_i$-inflated. This result is observed in Figure~\ref{f:CountSims} where $q_{X_A} = 0\cdot2$, $q_{X_B} = 0\cdot4$ and $\lambda_{X_A}$, $\lambda_{X_B} = 10$. Consider fitting an incorrectly specified model with $X_{A_i}|N_i \sim \mbox{binomial}(N_i, p)$. The true proportion $p$ of species A at a given site is 0.5, however the maximum likelihood estimate  is $0\cdot55$ with a 95\% confidence interval of $[0\cdot53, 0\cdot57]$. For the ZIB model of Hall the estimates are $0\cdot60$ (95\% CI $[0\cdot58, 0\cdot64])$. In both cases the estimates of $p$ are biased upwards as the model does not account for the spurious sources of $N$'s in the data. For species B the opposite case of underestimation of the species proportion will be the result. \cite{Royle2006} display simulations for imperfect detection in the Bernoulli setting which provide weight to this conclusion.  


For the pollen reconstruction problem the presence of zero \& $N$-inflation is relatively simple to explain, or at least tallies with the generating mechanism described above - the counts of Pinus D. and Juniper are individually zero-inflated for a variety of reasons. These include observer error in identification of pollen spores, or the absence of information on other important climate features. An alternative count generating mechanism comes from considering the counts as naturally arising from a mixture distribution, where the mixture reflects underlying structure within the experiment. For example, Hall's formulation of the ZIB model is based on the understanding of two underlying processes simultaneously generating counts which reflect either full or partial effectiveness in applications of the pesticide. Intuitively, this mixture framework can be extended to include processes which reflect additional count generating mechanisms. For example, in the whitefly example, a plausible biological explanation is that an individual treatment can potentially involve three outcomes - (1) the treatment is fully effective or unrecorded factors, for example the initial vitality of the whiteflies, result in zero survival, (2) the treatment is partially effective resulting in some whitefly deaths, and (3) the treatment is fully ineffective, resulting in full survival. Next, we present a mixture distribution model which reflects this intuition.


\section{An alternative generating mechanism for zero \& $N$-inflated binomial counts}\label{derivation}\label{ZNIB_APP}

Here we propose an alternative generating mechanism for zero/$N$ inflated counts which fits more within the biological explanation of such experiments. Suppose that  $Y_1+Y_2=N$, where both $Y_1$ and $Y_2$ are counts which independently arise from two separate zero-inflated Poisson processes \citep{Lambert}, i.e. $Y_1 \sim \mathrm{ZIP}(\mu_{Y_1},q_{Y_1})$ and $Y_2 \sim \mathrm{ZIP}(\mu_{Y_2},q_{Y_2})$ such that:

\vspace{5mm}

\begin{minipage}{0.45\textwidth}
$$ Y_1 \sim \left\{ \begin{array}{lcc} 0 & \pi_0 = 1-q_{Y_1} \\ 
\mathrm{Poisson}(\mu_{Y_1}) &  \pi_{\neq0} = q_{Y_1}  
 \end{array} \right.$$
\end{minipage}%
\hfill
\begin{minipage}{0.45\textwidth}
$$ Y_2 \sim \left\{ \begin {array}{lcc} 0  &  \pi_0 = (1-q_{Y_2}) \\ 
\mathrm{Poisson}(\mu_{Y_2}) &  \pi_{\neq0} =q_{Y_2}  
 \end{array} \right.$$ 
 \end{minipage}

\vspace{5mm}

\noindent
Now $Y_1 +Y_2 = N$ is a mixture containing four different components:

$$ N \sim \left\{ \begin{array}{lcc} 0 & \mbox{with probability} &  (1-q_{Y_1})(1-q_{Y_2}) \\ 
\mathrm{Poisson}(\mu_{Y_1}) & \mbox{with probability} & (1-q_{Y_2})q_{Y_1} \\ 
\mathrm{Poisson}(\mu_{Y_2}) & \mbox{with probability} & (1-q_{Y_1})q_{Y_2} \\ 
\mathrm{Poisson}(\mu_{Y_1}+\mu_{Y_2}) & \mbox{with probability} & q_{Y_1} q_{Y_2} \\ 
 \end{array} \right.$$ 

The conditional distribution $Y_1|N$ is now considered for all possible options. First, for $Y_1=0|N=0$, $\pi_{Y_1|N}(Y_1=0|N=0) = 1$. Now, suppose that $p =\mu_{Y_1}/(\mu_{Y_1} + \mu_{Y_2})$, namely that the probability parameter for success, $p$, in an $N$ constrained binomial trial, is the rate parameter of the $Y_1$ process divided by the sum of the rate parameters for the $Y_1$ \& $Y_2$ processes. Then for $Y_1=0$ and $N>0$:

\begin{eqnarray*}
\pi_{Y_1|N}(Y_1=0|N) &=& \frac{ ( 1-q_{Y_1})q_{Y_2} e^{\mu_{Y_1}} (1-p)^{N} + q_{Y_1} q_{Y_2} (1-p)^{N} }{(1-q_{Y_1}) q_{Y_2} e^{\mu_{Y_1}} (1-p)^{N} + (1-q_{Y_2}) q_{Y_1} e^{\mu_{Y_2}} {p}^{N} + q_{Y_1} q_{Y_2} } \\
\end{eqnarray*}

The next case is when $N>0$ and $Y_1=N$:

\begin{eqnarray*}
\pi_{Y_1|N}(Y_1=N|N) &=& \frac{ ( 1-q_{Y_2})q_{Y_1} e^{\mu_{Y_2}} p^{N} + q_{Y_1} q_{Y_2} p^{N} }{(1-q_{Y_1}) q_{Y_2} e^{\mu_{Y_1}} (1-p)^{N} + (1-q_{Y_2}) q_{Y_1} e^{\mu_{Y_2}} p^{N} + q_{Y_1} q_{Y_2} } \\
\end{eqnarray*}

The final case is for $1<Y_1<N$ and $N>0$:

\begin{eqnarray*}
\pi_{Y_1|N}(Y_1|N) &=& {N \choose Y_1} \frac{ q_{Y_1} q_{Y_2} {p}^{Y_1} (1-p)^{N-Y_1} }{(1-q_{Y_1}) q_{Y_2} e^{\mu_{Y_1}} (1-p)^{N} + (1-q_{Y_2}) q_{Y_1} e^{\mu_{Y_2}} p^{N} + q_{Y_1} q_{Y_2} } \\
\end{eqnarray*}

Taken together we obtain a zero \& N inflated binomial (ZNIB) distribution:

\begin{eqnarray} 
Y_1 \sim \left\{ \begin{array}{lllc}
0 & & \mathrm{with} \  \mathrm{probability} & q_{0}  \\
N & & \mathrm{with} \  \mathrm{probability} & q_{N} \\

\mathrm{bin}(N,p) & & \mathrm{with} \  \mathrm{probability} &  1 - q_{0} - q_{N}  \end{array}
 \right. \label{eqappend2}
\end{eqnarray}

where:
$$q_{0} = \frac{ ( \frac{1-q_{Y_1}}{q_{Y_1}}) e^{\mu_{Y_1}} (1-p)^{N} }{(\frac{1-q_{Y_1}}{q_{Y_1}}) e^{\mu_{Y_1}} (1-p)^{N} + (\frac{1-q_{Y_2}}{q_{Y_2}} e^{\mu_{Y_2}} )p^{N} + 1 }, q_{N} = \frac{ ( \frac{1-q_{Y_2}}{q_{Y_2}}) e^{\mu_{Y_2}} p^{N} }{(\frac{1-q_{Y_1}}{q_{Y_1}}) e^{\mu_{Y_1}} (1-p)^{N} + (\frac{1-q_{Y_2}}{q_{Y_2}}) e^{\mu_{Y_2}} p^{N} + 1 }$$

A reparameterisation of $q_0$ and $q_N$ in terms of zero/$N$ inflation parameters $\eta_{0}$ and $\eta_{N}$ benefits the notation as well as simplifying inference procedures by imposing $q_{0}$ $+$ $q_{N}$ $\leq$ $1$ $\forall$ $\eta_0$, $\eta_N$ $\in$ $\Re$. 

$$q_{0} = \frac{e^{\eta_{0}}}{1+e^{\eta_{0}}+e^{\eta_{N}}},  q_{N} = \frac{e^{\eta_{N}}}{1+e^{\eta_{0}}+e^{\eta_{N}}}$$

A convenient reformulation of the model in (\ref{eqappend2}) is as a mixture of three binomial distributions \{bin$(N,0)$, bin$(N,1)$, bin$(N,p)$\} with weights ${\tau}$ = ($q_0, q_N, 1-q_{0}-q_{N})$ and $\mathbf{p} = (0, 1, p)$. The probability mass function can then be written as:

\begin{eqnarray}
\mathrm{pr}(Y=k| {\tau}, \mathbf{p}) = \sum_{j=1}^3 \tau_{j} \mathrm{pr}_j(Y=k|p_j) \label{BinMix} 
\end{eqnarray}

Here $\mathrm{pr}_j$ is the probability mass function of the binomial distribution with proportion $p_j$.

Following the notation of \citet{Hall} and \citet{Tian2015} we consider zero \& $N$ inflated observations $y_i$, with variable sum constraints $N_i$,  for $i = 1,\ldots,n$.  The probability of an individual observation being zero or $N_i$ inflated is $q_{0_i}$ and $q_{N_i}$ respectively:

\begin{eqnarray}
Y_i \sim \left\{ \begin{array}{lllc}
0 & & \mathrm{with} \  \mathrm{probability} & q_{0_i} \\
N_i & & \mathrm{with} \  \mathrm{probability} & q_{N_i} \\
\mathrm{bin}(N_i,p_i) & & \mathrm{with} \ \mathrm{probability} & 1 - q_{0_i} - q_{N_i} \end{array}
 \right. \label{eq1}
\end{eqnarray}

The likelihood form in Equation~\ref{eq1} naturally arises as the convolution of two zero-inflated Poissons conditional on their sum total, or alternatively as a weighted mixture of binomial likelihoods. Here $E[Y_i] = \mu_i  =  q_{N_i}N_i +(1-q_{0_i} -q_{N_i})N_ip_i$ and  $Var(Y_i) = q_{N_i}N_i^2 + (1-q_{0_i}-q_{N_i})(N_ip_i)(1-p_i +N_ip_i) - \mu_i^2$. Estimation of model parameters via optimisation is complicated by the constraint that $0 \leq q_{0_i}, q_{N_i} \leq 1$ and $q_{0_i}+q_{N_i} \leq 1$. A reparameterisation of $q_{0_i}$ and $q_{N_i}$ in terms of zero \& $N$ inflation parameters $\eta_{0_i}$, $\eta_{N_i}$ simplifies inference procedures by imposing $q_{0}$ $+$ $q_{N}$ $\leq$ $1$ $\forall$ $\eta_{0_i}$, $\eta_{N_i}$ $\in$ ${\rm I\!R}$. 
\begin{eqnarray}
q_{0_i} = \frac{e^{\eta_{0_i}}}{1+e^{\eta_{0_i}}+e^{\eta_{N_i}}}, \;\;\;  q_{N_i} = \frac{e^{\eta_{N_i}}}{1+e^{\eta_{0_i}}+e^{\eta_{N_i}}}
\label{eqQ}
\end{eqnarray}
Note, if $q_{N_i} = 0$ (or equivalently, $\eta_{N_i} \leq -\infty$)  then the model simplifies to the ZIB form of \citet{Hall}. 

The final distribution and parameterisation is equivalent to the form presented by \citet{Tian2015}, though having a substantially different generating mechanism. Maximum likelihood methods, including details of an EM algorithm for fitting the
models, are provided in \citet{Tian2015} while \cite{Dupuy} provides the
theoretical properties of ML estimation.

\subsection{Overdispersed zero \& $N$ inflated binomial models}\label{ZEIB_APP}

Here we sketch a novel extension of the ZNIB model for situations where the likelihood does not sufficiently capture the variability, additional to the excess of zeros and $N$'s, in the observed counts. The derivation falls along similar lines to the ZNIB in the previous section, and thus we suppress explicit details. Suppose that  $Y_1+Y_2=N$, where both $Y_1$ and $Y_2$ are counts which independently arise from two separate zero-inflated negative binomial processes with matching probability parameter $p$, i.e $ Y_1 \sim 0$ with probability $(1-q_{Y_1})$ and negative binomial$(r_1,p)$  with probability $q_{Y_1}$,  and $ Y_2 \sim 0$ with probability $(1-q_{Y_2})$ and negative binomial$(r_2,p)$  with probability $q_{Y_2}$.  Generically, if $Y_1 \sim NB(r_1,p)$ and $r$ is real, then $Pr(Y_1=k) = \frac{\Gamma(k+r_1)}{k! \Gamma(r_1)}p^p(1-p)^{r_1}$. In terms of the sum constraint:

\begin{eqnarray*}
Pr(Y_1=k | Y_1+Y_2 = N) &=& 
                     \frac{\Gamma(k+r_1)\Gamma(N-k+r_2)}{\Gamma(N+r_1+r_2)}
                            \frac{\Gamma(r_1+r_2)}{\Gamma(r_1)\Gamma(r_1)}
                            \frac{N!}{k!(N-k)!}\\
                           &\sim&\mathrm{beta-binomial}(N,r_1,r_2) 
\end{eqnarray*}

Thus, sum constrained negative binomial random variables with matching $p$ follow a beta-binomial distribution. Replacing the Poisson likelihoods in the steps outlined in Section~\ref{derivation} with negative binomial likelihoods with matching $p$, it is straightforward to show that two sum -constrained zero-inflated negative binomial distributed variables with matching $p$ follow a zero \& $N$-inflated beta-binomial distribution (ZNIBB). A parameterisation of this likelihood in terms of variable $N_i$, $p_i$ and $s$ is convenient for modelling purposes, where $s$ is a parameter governing the overdispersion present across both species, i.e. a parameter that inflates the variance in terms of the mean-variance relationship assumed by the underlying binomial model with $ \frac{s+N_i}{s+1} \times N_ip_i(1-p_i) $. Increasing values of $s$ indicate reduced overall overdispersion.

\begin{eqnarray}
Y_i \sim \left\{ \begin{array}{lllc}
0 & & \mathrm{with} \  \mathrm{probability} & q_{0_i} \\
N_i & & \mathrm{with} \  \mathrm{probability} & q_{N_i} \\
\mathrm{beta-binomial}(N_i,p_i, s) & & \mathrm{with} \ \mathrm{probability} & 1 - q_{0_i} - q_{N_i} \end{array}
 \right. \label{eq2}
\end{eqnarray}

This likelihood naturally arises as the convolution of two zero-inflated negative binomials conditional on their sum total, and represents an extension of the zero-inflated beta-binomial model introduced by \cite{Hall2002}. The model simplifies to their parameterisation when $q_{N_i} = 0$ (or equivalently, $\eta_{N_i} \leq -\infty$). The link between model covariates and the $q_{0_i}$, $q_{N_i}$ is the same as in Equation~\ref{eqQ}.

\subsubsection{Moments of the distribution}\label{Moments}

The reformulation of the likelihood as a mixture of beta-binomial components leads to a simple expression for the moments of the distribution.  Let $(\tau_1, \tau_2, \tau_3) = (q_0, q_N, q = 1-q_0 -q_N)$. The expected value of each beta-binomial component is  $\mu_0 = 0$, $\mu_N = N$,  and $\mu^{\prime} = \frac{N_ir_1}{r_1 + r_2}$ resulting in $\mu = qN[p + q_N/q]$. It follows that $\left(\mu_0 - \mu \right) = - \mu; \left(\mu^{\prime} - \mu \right) = N[p(1-q) - q_N]$, and $\left(\mu_N- \mu  \right)= N[(1-qp)-q_N]$. We note also that for the degenerate first and last  components all moments are zero. It follows that the $j^{\mathrm{th}}$ central moment for the distribution is:
$$
E\left[ (X-\mu)^j \right] = q_0 [-\mu]^j + q_N \left( N[(1-qp)-q_N] \right)^j
+q \sum_k {j \choose k} (Np - \mu)^{j-k} m^{(k)}
$$
where $m^{(k)}$ denotes the $k^{th}$ central moment for the binomial distribution. Here $E[Y] = \mu  =  q_{N}N +(1-q_0 -q_N)Np$ and  $Var(Y) = q_{N}N^2 + (1-q_{0}-q_{N})(Np)(1-p +Np) - E[Y]^2$. Further moments can be obtained as necessary.

\section{Applications} \label{Applications}

In Section~\ref{HallApp} we demonstrate the superior fit of a ZNIB model to the whitefly dataset of \cite{Hall}, and in Section~\ref{PollenApp} demonstrate the superior and more credible fit of a ZNIBB model to the pollen dataset of \cite{Huntley}.

\subsection{Whitefly dataset}\label{HallApp}
The 640 data points for the 54 plants, grouped into 18 trios, is organised as follows: $y_{jklt}$ is the number of surviving insects recorded on plant $j$ ($j$ = 1, \ldots, 3), in treatment $k$ ($k$ = 1, \ldots, 6), in block $l$ ($l$ = 1, \ldots, 3), and recorded at time $t$ ($t$ = 1, \ldots, 12). $y_{jklt}$ is a count between 0 and $N_{jklt}$, where $N_{jklt}$ is the number of whitefly alive in each experiment as it is initialised; $N_{jklt}$ ranges in value from $1-15$. Due to the apparent efficacy of treatments, in 339/640 experiments (53\%) there are no surviving insects ($y_{jklt}$  = 0). Conversely, in 76/640 (12\%) of the experiments there is zero mortality, i.e.  $y_{jklt}$ = $N_{jklt}$. As previously noted, a histogram of the proportion of surviving whiteflies in Figure~\ref{f:Hall} suggests substantial zero and $N$-inflation in treatment 5 (the control), as well as highlighting the poor predictive performance of the fixed effects ZIB model for this treatment. The instances of zero mortality ($N$-inflation) occur for the control treatment across several plants, and are equally split across blocks; there are no obvious patterns or trends in the zero mortality evident from an initial exploratory analysis of the data. 

\subsubsection{ZNIB extension of Hall (2000)}\label{fixedeffects}

We incorporate an $N$-inflation component into the best fitting fixed effects ZIB model identified by \cite{Hall}, and investigate specifications for the $N$-inflation which include main effects and interactions among the factors treatment, block, and week. In terms of notation, let $\boldsymbol{\theta_0} = \{\theta_{0_1}, \ldots, \theta_{0_n} \}^{T}$, $\boldsymbol{\theta_N} = \{\theta_{N_1}, \ldots, \theta_{N_n} \}^{T}$, and $\mathbf{p} = \{p_{1}, \ldots, p_{n} \}^{T}$. We model the zero, $N$ and probability components as:

\begin{eqnarray}
\boldsymbol{\theta_0}  = \mathbf{B}\boldsymbol{\beta_0}, \;\;\;\; \boldsymbol{\theta_N} = \mathbf{D}\boldsymbol{\beta_N}, \;\;\;\; \mbox{logit}(\mathbf{p})  =  \mathbf{G}\boldsymbol{\beta_p}
\end{eqnarray}


where $\mathbf{B}$, $\mathbf{D}$ and $\mathbf{G}$ are design matrices of chosen covariates and $\boldsymbol{\beta_0}$, $\boldsymbol{\beta_N}$ and $\boldsymbol{\beta_p}$ the regression parameters associated with the covariates in each of the zero-inflation, $N$-inflation and binomial probability models. We assume $y_{jklt} \sim \mbox{ZNIB} (N_{jklt}, q_{0_{klt}}, q_{N_{klt}}) $, and find the estimates of model parameters via numerical maximisation of the log likelihood $\ell$ using Newton-Raphson optimisation, where:

\begin{eqnarray}
\ell(\boldsymbol{\beta_0},\boldsymbol{\beta_N},\boldsymbol{\beta_p}; \boldsymbol{y},\boldsymbol{n}) 
&=& \sum_{i=1}^{640} \mbox{log} \left\{e^{\mathbf{B_i}\boldsymbol{\beta_0}}\mathbb{1}_{0} + e^{\mathbf{D_i}\boldsymbol{\beta_N}}\mathbb{1}_{N} +  {N_i \choose y_i}\frac{e^{y_i\mathbf{G_i}\boldsymbol{\beta_p}}}{(1+e^{\mathbf{G_i}\boldsymbol{\beta_p}})^{n_i}}\right\}  -  \mbox{log} (1 + e^{\mathbf{B_i}\boldsymbol{\beta_0}} + e^{\mathbf{D_i}\boldsymbol{\beta_N}})     \nonumber 
\end{eqnarray}

The maximum likelihood estimates for model parameters converge quickly for all models, across all combinations of variables - this tallies with the previous experiences of \cite{Hall} and \cite{Lambert} in using such methods. We compare model performance via the negative form of $BIC = \mbox{ln}(\hat{L}) - k\;\mbox{ln}(n)/2$ presented in \cite{Hall}.  Investigation of various specifications for $\theta_N$ reveals the model: 

$$ \mbox{logit}(p_{klt}) = \mu_p + \mbox{block}_k +  \mbox{trt}_l + \mbox{week}_t+ (\mbox{trt} \times \mbox{block})_{lk}  + (\mbox{trt} \times \mbox{week})_{lt} $$
with
$$ \theta_{_0klt} = \mu_{\theta_{0}} +  \mbox{block}_k + \mbox{trt}_l + \mbox{week}_t,\qquad  \theta_{_Nklt}  = \mu_{\theta_{N}} + \mbox{trt}_l $$

As most of $y=N$ experiments are observed for treatment 5, it is unsurprising that the best fitting model for $\theta_N$ contains a treatment effect. However, other covariate effects also impact on the $N$-inflation probabilities via the normalising constant of ($1 + e^{\mathbf{B}\boldsymbol{\beta_0}} + e^{\mathbf{D}\boldsymbol{\beta_N}}$), which includes the covariates involved in modelling the zero-inflation. 

The parameter estimates for many of the parameters shared by the ZNIB and ZIB models are broadly equivalent, though the fixed effect for treatment 5 for $p$ reduces from $3\cdot78$ in the ZIB model to $2\cdot38$. This suggests that the estimated efficacy of the  control treatment is higher when partially effective than suggested by the ZIB fit - this is due to many of the $y_{jklt}=N_{jklt}$ cases being captured by the $N$-inflation (completely ineffective) process. The small number of significant differences  between parameter estimates relate to some of the interaction effects, with several of the (treatment $\times$ week) effects reducing substantially in magnitude. This is perfects a reflection of model parsimony, with the interaction effects in the ZIB model previously attempting to account for the extra $N$'s. The ZNIB model is more computationally stable, as evidenced by an invertible hessian matrix at the MLE, which is unavailable for the ZIB fit of \cite{Hall}. This suggests that the model may be overparameterised, perhaps explaining Hall's identification of a significant disorderly interaction between week and the subirrigation treatments which inhibits his marginal comparison between the treatments. We defer further discussion of these issues and estimates of model parameters to Section~\ref{OtherModels}.

%
%
%
%

In terms of model fit: the ZNIB model yields a maximum log likelihood of $-782\cdot7$ on 533 residual degrees of freedom and a $BIC$ of $-1134\cdot8$. By comparison, the ZIB (i.e. Hall) version of this model gives a maximum log likelihood value of $-851\cdot6$ and a $BIC$ of $-1184\cdot4$, highlighting the substantial improvement in predictive performance. This is further observed in Table~\ref{Table1}: here the difference in observed and predicted counts reveals that the ZNIB model is a more likely generating model for the data ($H_0$: generating distribution is ZNIB)with a $p$ value of $0\cdot07$, compared to $p$ = $0\cdot02$ for the ZIB model, based on comparisons with a $\chi^2(9)$ distribution.

\begin{table}
 \centering
  \caption{Observed values and predictions for the number of trials with the proportion of live insects at trial end falling within fixed intervals} \label{7.tableHall}
  \begin{tabular*}{\textwidth}{@{}l@{\extracolsep{\fill}}r@{\extracolsep{\fill}}r@{\extracolsep{\fill}}r@{\extracolsep{\fill}}}
  \hline
 & & \multicolumn{1}{c}{{Model}} &\\ 
\cline{3-4} \\ [-8pt]
 $y_i/N_i$     &  Observed   & ZIB ($\frac{(O_i-E_i)^2}{E_i})$&  ZNIB ($\frac{(O_i-E_i)^2}{E_i})$ \\ 
 \hline
0 - 0.1 &  353   & $348\cdot4\phantom{1}$ ($0\cdot06$)& $348\cdot3$ ($0\cdot06$)  \\
 0.1 - 0.2 & 45 & $40\cdot9\phantom{1}$ ($0\cdot41$)& $41\cdot6$ ($0\cdot27$)\\
 0.2 - 0.3 & 69 & $56\cdot6\phantom{1}$ ($2\cdot72$) & $60\cdot5$ ($1\cdot21$)\\
 0.3 - 0.4 & 10 & $15\cdot4\phantom{1}$ ($1\cdot87$)& $16\cdot7$ ($2\cdot68$)\\
 0.4 - 0.5 & 17 & $24\cdot8\phantom{1}$ ($2\cdot48$)& $27\cdot7$ ($4\cdot11$)\\
  0.5 - 0.6 & 30 & $32\cdot5\phantom{1}$ ($0\cdot18$)& $35\cdot6$ ($0\cdot88$)\\
   0.6 - 0.7 & 15 & $17\cdot2\phantom{1}$ ($0\cdot29$)& $14\cdot3$ ($0\cdot04$)\\
    0.7 - 0.8 & 10 & $7\cdot9\phantom{1}$ ($0\cdot55$)& $4\cdot6$ ($6\cdot43$)\\
     0.8 - 0.9 & 8  & $\pmb{23\cdot3}$ $(10\cdot08)$ & $9\cdot1$ ($0\cdot13$) \\
     0.9 - 1 & 83 & $\pmb{73\cdot0}\phantom{1}$ $(1\cdot37)$& $81\cdot7$ ($0\cdot02$)\\
\hline
$\sum_{i=1}^{10} \frac{(O_i-E_i)^2}{E_i}$ & & ($20\cdot01$) & ($15\cdot83$) \\
\hline
\end{tabular*}
\label{Table1}
\end{table}



\begin{figure} \centerline{\includegraphics[width=6.5in]{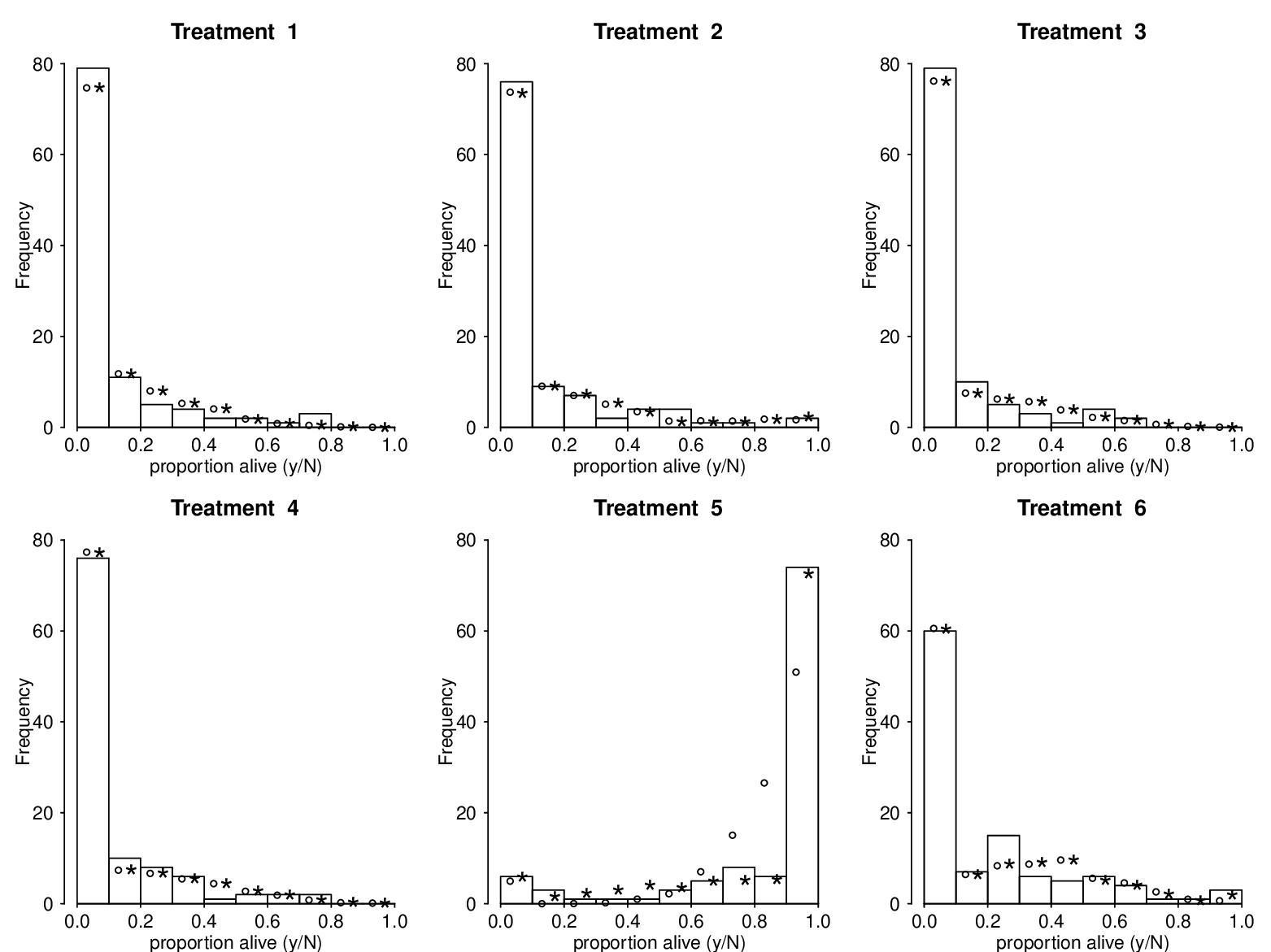}}
\caption{Frequency histogram of the surviving proportions ($y_i/N_i$) for the experiments within each treatment category. The predicted frequencies of the ZIB ($\circ$) and ZNIB ($\star$) models are overlain.}
\label{f:Hallboth} \end{figure}

In Figure~\ref{f:Hallboth} we observe that the predicted proportions of the ZNIB model reflect much better the observed data with a substantial improvement for treatment 5 in particular. The improvements for treatments 2 and 6 are more marginal, with the model indicating that the $N$'s observed in these cases are perhaps more reflective of natural variability. 

\subsubsection{Repeated measures}\label{RepeatedMeasures}

Given the series of repeated measures on each of the 54 plants over the 12 week period, \cite{Hall} presents a mixed model extension of the ZIB model which results in a superior $BIC$ of $-1175\cdot6$. Note this $BIC$ is substantially inferior to the ZNIB fit of the previous section, but for completeness we investigate the performance of a ZNIB mixed model fit. We account for plant to plant heterogeneity via the incorporation of a normal random effect $b_i\sim N(0,\sigma^2)$ for each plant in the specification of $\mathbf{p}$, and learn $\sigma$ from the data. Let $\mathbf{b}  = \{b_1, \ldots b_1, b_2, \ldots, b_2, \ldots b_{54}, \ldots, b_{54} \}$, denote the set where the random effect for each plant is replicated for each of the $12$ time points. This vector is corrected for the 8 missing observations, and is thus of length 640, matching the number of available datapoints. The log-likelihood for this mixed effects model is:

\begin{eqnarray}
\ell(\boldsymbol{\beta_0},\boldsymbol{\beta_N},\boldsymbol{\beta_p}, \sigma; \boldsymbol{y},\boldsymbol{n}) &=&
 \sum_{i=1}^{640} \int_{-\infty}^{\infty} \mbox{log} \left\{e^{\mathbf{B_i}\boldsymbol{\beta_0}}\mathbb{1}_{0} + e^{\mathbf{D_i}\boldsymbol{\beta_N}}\mathbb{1}_{N} +  {n_i \choose y_i}\frac{e^{y_i(\mathbf{G_i}\boldsymbol{\beta_p}+\sigma b_i )}}{(1+e^{\mathbf{G_i}\boldsymbol{\beta_p}})^{n_i}}\right\}db_i   \nonumber  \\
 && -  \mbox{log} (1 + e^{\mathbf{B_i}\boldsymbol{\beta_0}} + e^{\mathbf{D_i}\boldsymbol{\beta_N}})   
\end{eqnarray}

We integrate out the plant specific random effect at each step of the maximisation procedure using Gaussian quadrature with 20 points. 

In contrast with the results of \cite{Hall} the plant specific effect here is much reduced - for the comparable mixed effects ZIB model fit the estimate of $\sigma$ reduces from $0\cdot43$ to $0\cdot31$, reflecting that a substantial portion of the heterogeneity in the counts is captured by the $N$-inflation component of the model. Notably, as we observe in Table~\ref{Table2}, the inclusion of a plant specific random effect in the ZNIB model \textit{is discouraged by the $BIC$}. The maximised log-likelihood for the ZNIB mixed model increases from $-782\cdot7$ to $-779\cdot9$, however model performance is worse than the ZNIB fit once model complexity is accounted for - the $BIC$ decreases from $-1134\cdot8$ to $-1135\cdot3$. This appears to confirm the suspicion that the plant specific random effects are attempting to compensate for the $N$-inflation present in the data as a result of the model not allowing for an ineffective treatment. 
 
\begin{table}
 \centering
  \caption{Comparison of best-fitting models to whitefly dataset. Models with ``mixed'' contain random effects for the block effect, as per Hall.} \label{7.tableHall2}
  \begin{tabular*}{\textwidth}{@{}l@{\extracolsep{\fill}}r@{\extracolsep{\fill}}r@{\extracolsep{\fill}}r@{\extracolsep{\fill}}}
  \hline
Method     &  Log likelihood   & d.f. & BIC  \\ 
 \hline
Binomial & $-1105\cdot5$ & 556 & $-1376\cdot9$ \\
ZIB & $-851\cdot6$ & 537 & $-1184\cdot4$\\
\mbox{ZIB} mixed & $-839\cdot6$ & 536 & $-1175\cdot6$  \\
\textbf{ZNIB} & $\pmb{-782\cdot7}$ & $\pmb{531}$ & $\pmb{-1134\cdot8}$\\
\mbox{ZNIB} mixed & $-779\cdot9$ &530 & $-1135\cdot3$ \\
\hline
\end{tabular*}
\label{Table2}
\end{table}


\subsubsection{Alternate model specifications}\label{OtherModels}
Whilst the best model identified by \cite{Hall}  includes main and interaction effects between the covariates, we have found that a much simpler model is perhaps to be preferred when penalising for model complexity.

$$ \mbox{logit}(p_{klt}) = \mu_p + \mbox{trt}_l  $$
with
$$ \theta_{_0klt} = \mu_{\theta_{0}} +  \mbox{trt}_l ,\qquad  \theta_{_Nklt}  = \mu_{\theta_{N}} + \mbox{trt}_l $$

Using only treatment as a covariate, the $BIC$ for the fixed effects ZIB model is $-1049\cdot95$, which reduces to $-1029\cdot38$ for a mixed model with plant specific random effects. This compares to $-974\cdot84$ for a fixed effects ZNIB fit, and $-959\cdot16$ for a mixed model ZNIB fit, which is identified as the best fitting model overall. \cite{Hall} does not explore this simpler model.

While the $BIC$ of the best model here, the ZNIB mixed model, is substantially lower than those in Table~\ref{7.tableHall2}, the conclusions in \cite{Hall} are reaffirmed. Active treatments with longer delays between last watering and application of the pesticide appear best at suppressing whitefly reproduction, and are most effective when subirrigation is used to deliver the pesticide. For example, treatment 4 - the application of the pesticide via subirrigation after 4 days without water, has a predicted full mortality of $0\cdot65$ (95\% approximate CI using the delta method $[0\cdot56,0\cdot75]$). When full mortality is not attained (i.e. only partially effective), the predicted survival under this treatment is $0\cdot30$ (95\% CI $[0\cdot20, 0\cdot42]$). Treatment 1 has the highest partial effectiveness, with a survival rate of  $0\cdot22$ (95\% CI $[0\cdot14, 0\cdot32]$), but the estimated full mortality is lower, indicating that the pesticide is more effective the longer plants go without water.

The primary differences between models relate to the control treatment. The control is predicted to be completely ineffective with probability $0\cdot63$ (95\% CI $[0\cdot54, 0\cdot74]$). There is no zero-inflation detected for the control (in contrast to the ZIB fit) and the predicted survival when partially effective is $0\cdot68$ (95\% CI $[0\cdot56, 0\cdot77]$). This may reflect experimental factors outside study control. This is perhaps also the case for the minor amounts of $N$-inflation detected in treatment 2 (95\% CI $[0\cdot005,0\cdot07]$), and 6 (95\% CI $[0\cdot004,0\cdot08]$). The estimate of $\sigma$ is $0\cdot39$, as compared to $0\cdot31$ for the model with interaction effects in Section~\ref{fixedeffects}, reflecting that the simple model still does a good job of capturing most of the variability in the data.

To summarise, the best fitting models overall are the $N$-inflated models, which reflect that there is evidence for several experimental units exhibiting $N$-inflation, or complete ineffectiveness at times of individual treatment applications. This can be attributed to factors outside of study design, such as errors in the delivery of the pesticide for treatments. The partial effectiveness of the control treatment can perhaps also be attributed to factors such as the initial vitality of the whiteflies, or for the simplest model, plant specific effects. These models compare favourably to those of \cite{Hall}, which display signs of overparameterisation.

\subsection{Palaeoclimate reconstruction using pollen data}\label{PollenApp}

The pollen dataset consists of counts for Juniper and Pinus D. at $4619$ sites in North America, with $MTCO$ (as before, a measure of the coldness of winter), also available at each site. Our primary interest is the construction of a model relating pollen abundance of Juniper to local $MTCO$, in order to quantitatively understand the preferred climate range of Juniper. A further interest is in using the fitted model to predict the $MTCO$ values of fossil pollen samples at sites where this information is unknown - \cite{Haslett} describe this as an ``inverse problem'', where the fitted model is inverted to make inferences on the missing covariate information. For illustrative purposes we randomly split the dataset 80:20 into a training dataset of $3695$ sites used for model fitting, and a validation set of 924 sites used for assessing the predictive accuracy of fitted models.


\subsubsection{Model \& Inference} \label{Inference}
Let $\mathbf{Y} = \{y_1, \ldots, y_{3695}\}$, represent the counts of Juniper, $\mathbf{Z} = \{z_1, \ldots, z_{3695}\}$ the counts of Pinus D, with sum constraints $\mathbf{N} = \{N_1, \ldots, N_{3695}\} = \mathbf{Y} +\mathbf{Z}$.  As previously mentioned, the $N_i$ are variable due to the differing number of pollen grains of each species counted at each site, ranging in value from 1 to 1000. Relative abundance of the species is important here as the counts of each are species are constrained by the sum total collected.

Due to the variability additional to the zero and $N$-inflation observed in the Juniper proportions in Figure~\ref{f:Pollen}, we specify a zero and $N$-inflated beta-binomial distribution for the counts, i.e. $\pi(y_i) \sim \mbox{ZNIBB}(y_i, N_i, p_i, s, q_{0_i}, q_{N_i})$, where $s$ describes the degree of overdispersion of the beta-binomial proportion of the model. We organise the data as follows - let $\mathbf{p} = (p_1, \ldots, p_{3695})^{T}$, $\boldsymbol{\theta_0} = (\theta_{0_1}, \ldots, \theta_{0_{3695}})^{T}$, $\boldsymbol{\theta_0} = (\theta_{N_1}, \ldots, \theta_{N_{3695}})^{T}$, with the $q_{0_i}$, $q_{N_i}$ functions of the $\theta$'s as outlined in Equation~\ref{eqQ}. 

We use Bayesian inference procedures due to their compatibility with the nature of the problem. Our ultimate aim is to use the model \textit{inversely} - the first stage of the problem involves the construction of a model relating pollen abundance ($\mathbf{Y}$, $\mathbf{N}$) to climate, i.e. $\pi(\mathbf{Y}, \mathbf{N}| MTCO)$, and then using the posterior samples from the first stage to sample from the posterior distribution of unknown $MTCO$ corresponding to new data, $(y^*, N^*)$, i.e. $\pi(MTCO | y^*, N^*, \mathbf{Y}, \mathbf{N})$, for which the $MTCO$ is unknown.

In terms of linking pollen abundance to climate, we expect the Juniper response to climate to smoothly vary in a non-monotonic manner - Juniper may have multiple preferred $MTCO$ ranges, as the pollen samples are comprised of several species of the same genus including \textit{Juniperus Communis}, \textit{Juniperus horizontalis} etc. As a result the Juniper response to climate is potentially multimodal, including those of the zero and $N$-inflation processes. To allow for non-linear flexibility in each of these relationships, we utilise a penalised spline model with B-spline cubic basis functions constructed from piecewise polynomial functions that are differentiable to degree 3~\citep{Eilers1996,Eilers2015}. We assign 35 spline knot points which are equally spaced across the range of $MTCO$ of $[-36^{\circ}C, 20^{\circ}C]$. This number of knots was arrived at by using cross validation to minimise curve overfitting whilst not oversmoothing the data.

\begin{eqnarray}
\mbox{logit}(\mathbf{p})  =  \mathbf{B}\boldsymbol{\beta_p}, \;\;\;\;
\boldsymbol{\theta_{0}}  =  \mathbf{B}\boldsymbol{\beta_0}, \;\;\;\;
\boldsymbol{\theta_{N}}  =  \mathbf{B}\boldsymbol{\beta_N} 
\end{eqnarray}

$\mathbf{B}$ is a $3695 \times 35$ of basis functions for the 3696 samples at the 35 knots, and $\boldsymbol{\beta_p} = (\beta_{p_1}, \ldots, \beta_{p_{35}})^T $, $\boldsymbol{\beta_N}$, $\boldsymbol{\beta_0}$ the B-spline coefficients with dimension $35 \times 1$. We impose a second order penalty on the basis coefficients, $\Delta^2(\beta_{p_k}) = \beta_{p_k}-2\beta_{p_{k-1}} + \beta_{p_{k-2}}  \sim N(0, \sigma_p^2)$, with uninformative priors specified for $\beta_{p_1}$ and $\beta_{p_2}$ $\sim \mathcal{N}(0, 3^2)$. For the zero-inflation and $N$-inflation processes we allow more flexibility in the response curves by imposing first order penalties, $\Delta(\beta_{0_k}) =\beta_{0_k}-\beta_{0_{k-1}} \sim \mathcal{N}(0, \sigma_0^2)$, $\Delta(\beta_{N_k}) \sim \mathcal{N}(0, \sigma_N^2)$, with $\beta_{0_1}$ and $\beta_{N_1}$ $\sim \mathcal{N}(0, 3^2)$. The $\sigma$'s control how closely related the basis coefficients are and will therefore control smoothness. 
We assign the following prior distributions on model parameters:
\begin{eqnarray}
\pi(\beta_{p_k}) &\sim& \mathcal{N}(2\beta_{p_{k-1}} - \beta_{p_{k-2}}, \sigma_p^2) \nonumber \\
\pi(\beta_{0_k}) &\sim& \mathcal{N}(\beta_{0_{k-1}}, \sigma_0^2) \nonumber \\
\pi(\beta_{N_k}) &\sim& \mathcal{N}(\beta_{N_{k-1}}, \sigma_N^2) \nonumber \\
\pi(\sigma_p) &\sim& \mathcal{N}(0\cdot5, 0\cdot1^2) \nonumber  \\
\pi(\sigma_0), 
\pi(\sigma_N),
\pi(s) &\sim& \mathcal{C}(0, 2\cdot5) 
\end{eqnarray}

We specify non-informative Cauchy priors on $s$, $\sigma_0$ and $\sigma_N$, allowing them to be learned from the data. Experimentation with prior elicitation for $\sigma_p$ reveals that a non-informative Cauchy prior results in difficulty separating the effects of Juniper response and zero-inflation probabilities at extremely cold values of $MTCO$  -  imposing an informative $\mathcal{N}(0\cdot5, 0\cdot1^2)$ prior for $\sigma_p$, resolves this issue. Biologically, a stronger prior on small $\sigma_p$ is more reasonable, as the pollen response to climate should be smoother than any zero or $N$-inflation features occurring due to local factors.

We use the RStan package \citep{rstan} within R, which uses Hamiltonian Monte Carlo methods to provide samples from the posterior distribution. We initialise 4 parallel chains, with a burn in of 250 samples for each chain to tune sampling parameters. We generate 250 samples from the posterior for each chain after burnin, which provides 1000 in total. Exploration of the convergence diagnostics indicate that no issues arise during sampling, and the posterior distribution  appears to be well explored - the $\hat{R}$ statistic measuring  the ratio of the average variance of samples within each chain to the variance of the pooled samples across chains is close to 1. Lag correlations for sequential samples from the posterior are also close to 0, highlighting the efficiency of the method in obtaining posterior samples. 


There are only minor differences between the 95\% highest posterior density (HPD) intervals for $\sigma_p$ and $\sigma_0$  of the ZNIBB model as compared to a zero-inflated only ZIBB model. A major difference can be observed however in comparing summaries of the overdispersion parameter $s$, with a 95\% HPD for $s$ of $[2\cdot8, 3\cdot4]$ for the ZNIBB model as compared to $[1\cdot9, 2\cdot4]$ for the ZIBB. Higher values of $s$ for the beta-binomial distribution actually indicate a \textit{decrease} in overdispersion for the ZNIBB model fit, indicating that the excess variability brought about by $N$-inflation is absorbed by $s$ within the ZIBB fit to the data.

\begin{figure}
 \centerline{\includegraphics[scale=.5]{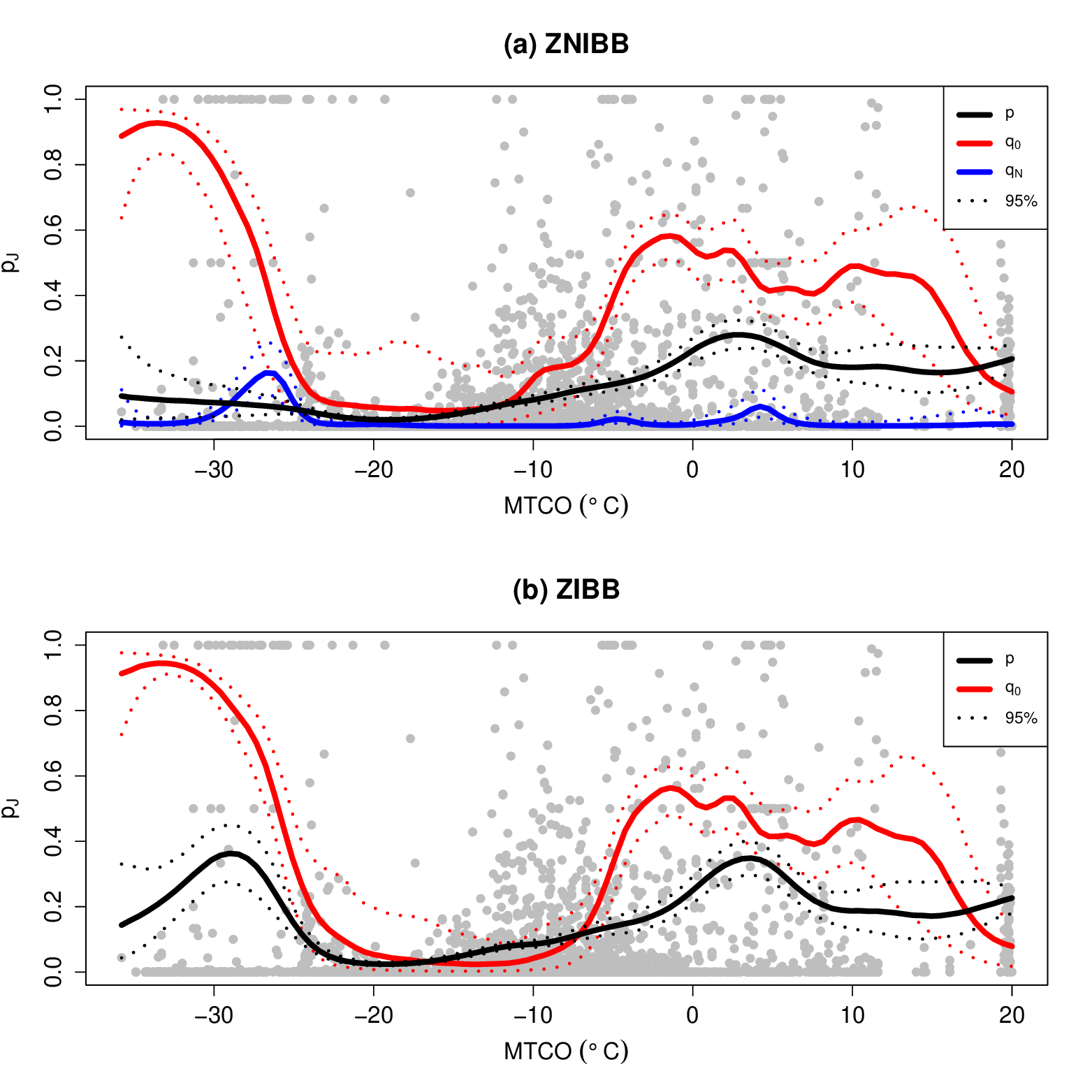}}
\caption{ Observed Juniper pollen proportions (\textcolor{mygray}{$\bullet$}). (a) \textbf{ZNIBB model:} fitted beta-binomial response probabilities, \textcolor{red}{zero-inflation} probabilities and \textcolor{blue}{$N$-inflation} probabilities with 95\% HPD bounds. (b) \textbf{ ZIBB model:} fitted beta-binomial probabilities and \textcolor{red}{zero-inflation} probabilities with 95\% HPD bounds.}
\label{f:Pollen_results}
\end{figure}

In Figure~\ref{f:Pollen_results} we project the basis functions to 100 equally spaced points on $[-36^{\circ}C, 20^{\circ}C]$, and present 95\% HPD's at each point for predictions of $p$, $q_0$ \& $q_N$.



There  is an extreme difference between the models in inferences on the preferred $MTCO$ range of Juniper. The $N$-inflation observed for Juniper at $MTCO$'s between $(-35^{\circ}C, -18^{\circ}C)$ in Figure~\ref{f:Pollen_results} (b) results in an overestimation of the overdispersion within the data for the ZIBB model. It also results in an implausible overestimation of $p$, the expected proportion of Juniper counts observed in that range. The fitted probabilities for the ZIBB model are not credible, the Juniper genus prefers warmer climates typically, and whilst its growing range can extend to colder regions than that of Pinus D., it would not be expected to comprise such a substantial proportion of the pollen assemblage at extremely low $MTCO$ values. This is due to the dispersal strategy of their respective pollen - the pollen of Pinus D. can be transported thousands of km from the source by wind due to the structure of the seed which facilitates long distance dispersal \citep{Campbell}. Juniper pollen, typically dispersed in fruit form, is more locally constrained. 

In light of this intuition, the results for the ZNIBB model presented in Figure~\ref{f:Pollen_results} (a) are much more credible - based on the pollen dispersal strategy of each genus we expect that the pollen of Pinus D. should dominate the assemblage at low $MTCO$ values which neither genus particularly favours. This is in contrast to the inferences of the ZIBB model fit. The $N$-inflation probabilities in Figure~\ref{f:Pollen_results} (a) also have three distinct peaks which indicate that the $N$'s observed at these $MTCO$ values are due to the unexpected absence of Pinus D. pollen (i.e. zero-inflation of the Pinus D. counts) as opposed to pollen of the Juniper genus naturally dominating the assemblage. 

\subsubsection{Climate reconstructions}\label{reconstructions}




We evaluate the predictive accuracy of the fitted models by inverting the fitted model to make inferences on $MTCO$ for the $924$ pollen in the left-out validation set for which the true $MTCO$ is known. In order to evaluate the posterior distribution for $MTCO$ given new data $(y^*, N^*)$, we discretise the $MTCO$ domain ($-36^{\circ}C$, $20^{\circ}C$) to 100 equally spaced grid points, $\mathbf{C} = (c_1,\ldots,c_{100})$. In the following we focus on the results, explicit details on the steps involved in model inversion and the construction of highest posterior density intervals are provided in the Appendix. 
In Figure~\ref{f:Pollen_recon1} we present 95\% highest posterior density intervals for a number of combinations of $(y^*, N^*)$ in the left-out validation set. We observe in each reconstruction that the ZIBB model typically places too much probability mass at cold regions of $MTCO$ due to the overestimation of the pollen response for Juniper at these points. Conversely, the ZNIBB model appears to perform much better in this regard, minimising the amount of unrealistic probability mass placed in extremely cold regions.

\begin{figure}
 \centerline{\includegraphics[scale =1]{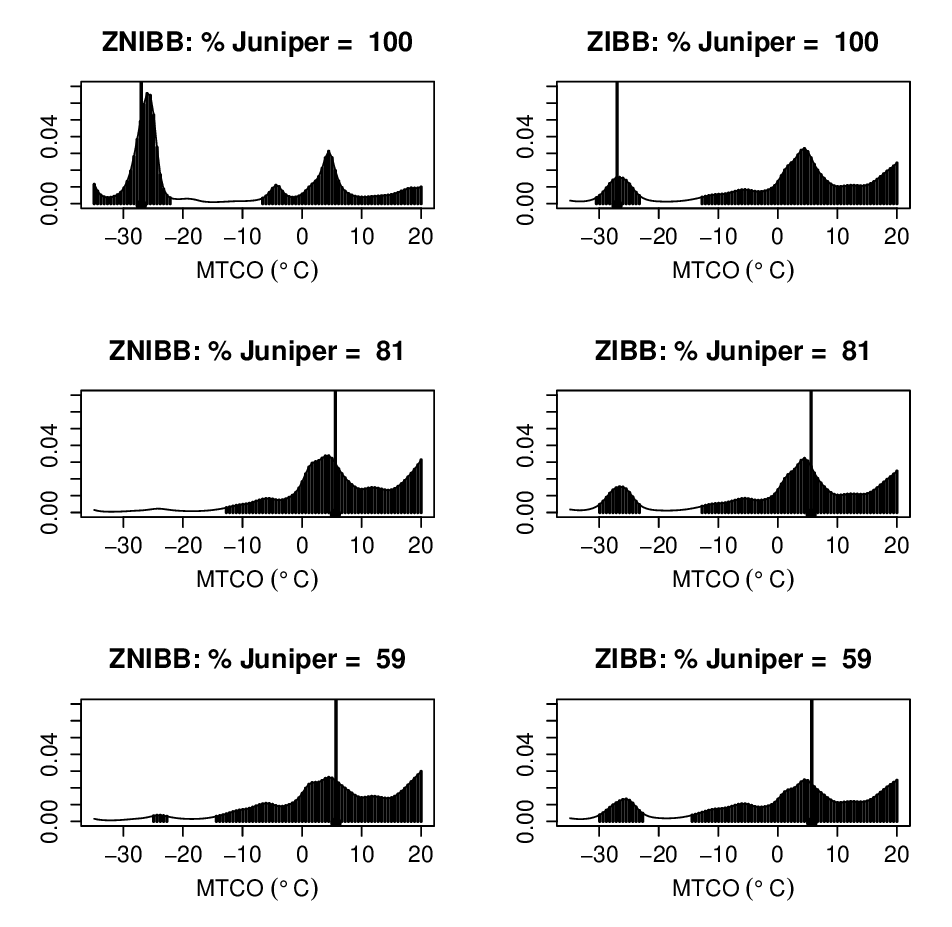}}
\caption{ 95\% HPD regions of $MTCO$ predicted by ZNIBB and ZIBB models across different Juniper proportions of the pollen record. Vertical line is the true $MTCO$ value which is known for the left out set.}
\label{f:Pollen_recon1}
\end{figure}

In Table~\ref{Table:Pollen} we assess predictive accuracy of the models via the Root Mean Squared Error of Prediction ($RMSEP$) and the \% of observations laying within their 50\%, 75\% and 95\% intervals. The $RMSEP$ provides a measure for assessing how close the posterior predictive intervals are to the known $MTCO$ locations within the left out set. If $y^*_i$, $N^*_i$ are the counts corresponding to left out datapoint $i$, and $MTCO_i$ the known $MTCO$ value, we evaluate the $RMSEP$ as: $$RMSEP (^{\circ}C) = \sqrt{\frac{\sum_{i=1}^{924}{\sum_{k=1}^{100}{(MTCO_i - c_{ik})^2\pi(c_{ik}|y^*_i,N^*_i,\mathbf{Y},\mathbf{N})}}}{924}}$$

When we separate the $RMSEP$ by strata In Table~\ref{Table:Pollen}, we observe that the performance of the ZNIBB is superior in most instances. For observations where $y=0$, the predictive accuracy of the models in terms of $RMSEP$ is approximately the same. However for counts between 0 and $N$, the ZNIBB model is $\approx 0\cdot4^{\circ}C$ more accurate on average ($17\cdot66^{\circ}C$ versus $18\cdot01^{\circ}C$), which represents a 4\% reduction in the Mean Squared Error of Prediction (MSEP). Furthermore, for counts where $y=N$ the predictive performance of the ZNIBB model is $\approx5^{\circ}C$ degrees superior ($23\cdot07^{\circ}C$ versus $28\cdot96^{\circ}C$),  which represents a substantial $\approx37\%$ reduction in the MSEP. This is in spite of the 95\% HPD intervals for the ZNIBB model appearing overly conservative (100\% coverage) for the $N$-inflated counts, which indicates that the posteriors are better centered on the true $MTCO$ for this model.

\begin{table}
 \centering
  \caption{Predictive performance of trained models on the left out observations set.} \label{tablePollen}
  \begin{tabular*}{\textwidth}{@{}l@{\extracolsep{\fill}}c@{\extracolsep{\fill}} r@{\extracolsep{\fill}}c@{\extracolsep{\fill}}c@{\extracolsep{\fill}}c@{\extracolsep{\fill}}c@{\extracolsep{\fill}}}
  \hline
Set & Model     &  \% within:& 50\% HPD &75\% HPD &95\% HPD  & RMSE ($^{\circ} C$)  \\ 
 \hline
$\mathbf{y = 0}$ &&&&&  \\
&ZIBB & &54 & $69$ &$97$ & $19\cdot 99 $ \\
&ZNIBB & &$\mathbf{53}$ & 69 &$97$ & $\mathbf{19\cdot 97}$ \\
\\
$\mathbf{0< y < N}$ &&&&&  \\
&ZIBB & &62 & $79$ &$98$ & $18\cdot 01 $ \\
&ZNIBB & &$\mathbf{58}$ & $\mathbf{78}$ &$98$ & $\mathbf{17\cdot66}$ \\
\\
$\mathbf{y = N}$ &&&&&  \\
&ZIBB & &36 & $55$ &$82$ & $28\cdot 96 $ \\
&ZNIBB & &$36$ &$ \mathbf{64}$ &$100$ & $\mathbf{23\cdot07}$ \\
\hline
\end{tabular*}
\label{Table:Pollen}
\end{table}

A further measure of interest is the \% of observations lying within the $(1-\alpha)$\% HPD regions, which should approximately equal  $(1-\alpha)$\%. We observe that the HPD regions are too conservative for both the ZNIBB and ZIBB models, however the performance of the ZNIBB model is superior within most categories.

\section{Summary}\label{Discussion}

The two applications considered in Section~\ref{Applications} illustrate the substantial impact that simultaneous zero and $N$-inflation can have on the inferences derived from upper bounded count data, and result in the re-evaluation of results from previously published studies. In the whitefly case of \cite{Hall}, ZIB models provide poor fits to some of the experimental treatments, and result in within-plant variation being erroneously flagged as substantially impacting on survival outcomes for the overparameterised model presented in \cite{Hall}. The  zero and $N$-inflation present in the counts of the control group results in the efficacy of this treatment being underestimated when it is partially effective, which is presumably due to factors outside of study design - the survival proportions of the whitefly are substantially overestimated. The ZNIB models in this case provides a natural explanation for the variation observed - survival counts may be zero-inflated due to full effectiveness of a treatment, or non-zero due to the partial effectiveness of a treatment. Furthermore, survival counts may be $N$-inflated due to treatment ineffectiveness, perhaps reflective of errors in the experimental process in terms of the delivery of treatments, or unaccounted-for factors impacting during the study. 

The pollen example fits with the interpretation of zero and $N$-inflated binomial counts naturally arising in the collection of zero-inflated Poisson counts subject to an upper bound. A small number of zeros (1.5\%) for the dominant Pinus D. genus result in misleading and implausible inferences by ZIB models on the pollen production response to climate. These misleading inferences are propagated and compounded when ZIB models are used for prediction of the unknown climates of fossil pollen, as is clearly illustrated in Section~\ref{PollenApp}. The superiority of the ZNIBB model is evidenced by the more coherent and plausible output of the fitted models as well as in the improved predictive performance. Indeed the methods in this paper have most obvious application in the area of palaeoclimate reconstruction - \cite{MST} have applied ZIBB models to pollen data for climate reconstructions which provides an obvious application for extension. Alternate climate proxies to pollen including chironomids and floramnifera may also be considered. These proxies, similarly used to provide the basis for estimates of past climatic conditions, are subject to the same problems of zero and $N$-inflation and we expect the improved inferences observed in this paper to transfer to these methods.



\section{Acknowledgement}\label{ack}
\noindent
Support of Science Foundation Ireland [11/PI/1027], Irish Research Council (New Foundations Award), and the Royal Irish Academy (Charlemont Award) is gratefully acknowledged. Further thanks to Professor Brian Huntley for his assistance in interpreting the biology of the pollen example, and to Professor Daniel B. Hall for kindly providing the whitefly dataset. 

\appendix

\section{}
We obtain samples of $\pi(p_{0_k}, q_{0_k}, p_{N_k} | \mathbf{Y},\mathbf{N})$ corresponding to each of the $(c_1,\ldots,c_k, \ldots,c_{100})$  via the 1000 posterior samples of $\pi(\boldsymbol{\beta_p}, \boldsymbol{\beta_0}, \boldsymbol{\beta_N},s | \mathbf{Y},\mathbf{N})$ from the model training stage outlined in Section~\ref{Inference}. To simplify model inversion, we separately evaluate the posterior probability at each of the $c_k$,  i.e. $\pi(c_k|y,n,\mathbf{Y},\mathbf{N})$, which we assume depend only on the $p_k$, $q_{0_k}$, $q_{N_k}$ at that point, as well as $s$:
\begin{eqnarray}
\pi(c_k | y^*, N^*, \mathbf{Y}, \mathbf{N}) &=& \int{\pi(c_k,p_k, q_{0_k},q_{N_k}, s| y^*, N^*, \mathbf{Y}, \mathbf{N})}dp_k \:dq_{0_k}\: dq_{N_k}\: ds\nonumber \\
&=& \int{\pi(c_k| p_k, q_{0_k},q_{N_k}, s) \pi(p_k, q_{0_k},q_{N_k}, s| \mathbf{Y}, \mathbf{N})}dp_k \:dq_{0_k}\: dq_{N_k}\: ds 
\label{eqnBayes}
\end{eqnarray}
As $y^*$ and $N^*$ are not used in model training, $\pi(p_k, q_{0_k},q_{N_k}, s| y^*, N^*, \mathbf{Y}, \mathbf{N})$ simplifies to $\pi(p_k, q_{0_k},q_{N_k}, s| \mathbf{Y}, \mathbf{N})$. Using Bayes' Theorem we rewrite Equation~\ref{eqnBayes} as:
\begin{eqnarray}
\pi(c_k | y^*, n^*, \mathbf{Y}, \mathbf{N}) 
&\propto& \int{\pi(y^*, n^* | p_k, q_{0_k},q_{N_k}, s) \pi(p_k, q_{0_k},q_{N_k}, s|\mathbf{Y}, \mathbf{N})\pi(c_k)}dp_k dq_{0_k} dq_{N_k} ds \nonumber
\end{eqnarray}

If we assume a flat prior on climate, $\pi(c_k) \propto 1$, then the unnormalised $\pi(c_k | y^*, N^*, \mathbf{Y}, \mathbf{N})$ can be evaluated via the posterior samples of $\pi(p_k, q_{0_k},q_{N_k}, s|\mathbf{Y}, \mathbf{N})$ as: $$\pi(c_k | y^*, N^*, \mathbf{Y}, \mathbf{N}) \approx \frac{1}{1000}\sum_{j=1}^{1000}{\pi(y^*, N^* | p_{k_j}, q_{0_{k_j}},q_{N_{k_j}}, s_j) }$$

The normalising constant for $\pi(c_1,\ldots,c_{100}| y^*, N^*, \mathbf{Y}, \mathbf{N})$ is found by summing the unnormalised posteriors of each $c_k$ across all 100 evaluation points $\sum_{k=1}^{100}\pi(c_k|y^*, N^*,\mathbf{Y},\mathbf{N})$. 

We obtain approximate $95\%$ highest posterior density regions for the predictions as follows:

\begin{enumerate}
\item Order the normalised $\pi(c_1,\ldots,c_{100}|y^*, N^*, \mathbf{Y}, \mathbf{N})$ from largest to smallest in posterior probability, obtaining $\pi(c^o_1,\ldots,c^o_{100}|y^*, N^*, \mathbf{Y}, \mathbf{N})$. Here $c^o_1$ represents the value of $c_k$ with largest posterior probability $\pi(c_k|y^*, N^*, \mathbf{Y}, \mathbf{N})$ and $\pi(c^o_1|y^*, N^*, \mathbf{Y}, \mathbf{N}) > \pi(c^o_2|y^*, N^*, \mathbf{Y}, \mathbf{N})\ldots$.
\item Initialise the empty set $\mathbf{S}$ and starting at $i=1$, add the $c^o_i$, until the cumulative sum of posterior probabilities for the members of $\mathbf{S}$ equals or just exceeds $0\cdot95$. 
\item The $c^o_i$ within $\mathbf{S}$, and their posterior probabilities, represent an approximate $95\%$ HPD for $MTCO$ given $y^*$, $N^*$ and the training data. 
\end{enumerate}

\bibliographystyle{apalike}
\bibliography{main}

\begin{thebibliography}{}

\bibitem[Bandyopadhyay et~al., 2011]{Bandyopadhyay}
Bandyopadhyay, D., Reich, B.~J., and Slate, E.~H. (2011).
\newblock A spatial beta-binomial model for clustered count data on dental caries.
\newblock {\em Statistical Methods in Medical Research}, 20(2):85--102.
\newblock PMID: 20511359.

\bibitem[Campbell et~al., 1999]{Campbell}
Campbell, I., McDonald, K., Flannigan, M.~M., and Kringayark, J. (1999).
\newblock Long-distance transport of pollen into the arctic.
\newblock {\em Nature}, 399:29--30.

\bibitem[Deng and Zhang, 2015]{Deng2015}
Deng, D. and Zhang, Y. (2015).
\newblock Score tests for both extra zeros and extra ones in binomial mixed regression models.
\newblock {\em Communications in Statistics - Theory and Methods}, 44(14):2881--2897.

\bibitem[Diallo et~al., 2019]{diallo2019estimation}
Diallo, A.~O., Diop, A., and Dupuy, J.-F. (2019).
\newblock Estimation in zero-inflated binomial regression with missing covariates.
\newblock {\em Statistics}, pages 1--27.

\bibitem[Dupuy, 2017]{Dupuy}
Dupuy, J.-F. (2017).
\newblock Inference in a generalized endpoint-inflated binomial regression model.
\newblock {\em Statistics}, 51(4):888--903.

\bibitem[Eilers et~al., 2015]{Eilers2015}
Eilers, P., Marx, B., and Durban, M. (2015).
\newblock Twenty years of p-splines.
\newblock {\em SORT-STATISTICS AND OPERATIONS RESEARCH TRANSACTIONS}, 39(2):149--186.

\bibitem[Eilers and Marx, 1996]{Eilers1996}
Eilers, P. H.~C. and Marx, B.~D. (1996).
\newblock {Flexible smoothing with B-splines and penalties}.
\newblock {\em Statistical Science}, 11(2):89 -- 121.

\bibitem[Guolo and Varin, 2014]{guolo2014}
Guolo, A. and Varin, C. (2014).
\newblock Beta regression for time series analysis of bounded data, with application to canada google ® flu trends.
\newblock {\em The Annals of Applied Statistics}, 8:74--88.

\bibitem[Hall, 2000]{Hall}
Hall, D.~B. (2000).
\newblock Zero-inflated poisson and binomial regression with random effects: A case study.
\newblock {\em Biometrics}, 56(4):1030--1039.

\bibitem[Hall and Berenhaut, 2002]{Hall2002}
Hall, D.~B. and Berenhaut, K.~S. (2002).
\newblock Score tests for heterogeneity and overdispersion in zero-inflated poisson and binomial regression models.
\newblock {\em Canadian Journal of Statistics}, 30(3):415--430.

\bibitem[Haslett et~al., 2006]{Haslett}
Haslett, J., Whiley, M., Bhattacharya, S., Salter-Townshend, M., Wilson, S.~P., Allen, J. R.~M., Huntley, B., and Mitchell, F. J.~G. (2006).
\newblock Bayesian palaeoclimate reconstruction.
\newblock {\em Journal of the Royal Statistical Society Series A: Statistics in Society}, 169(3):395--438.

\bibitem[Huntley, 1993]{Huntley}
Huntley, B. (1993).
\newblock The use of climate response surfaces to reconstruct palaeoclimate from quartenary pollen and plant microfossil data.
\newblock {\em Philosophical Transactions of the Royal Society of London B: Biological Sciences}, 341(1297):215--224.

\bibitem[Joseph et~al., 2016]{Joseph2016}
Joseph, M.~B., Preston, D.~L., and Johnson, P. T.~J. (2016).
\newblock Integrating occupancy models and structural equation models to understand species occurrence.
\newblock {\em Ecology}, 97(3):765--775.

\bibitem[Koslovsky, 2023]{Koslovsky2023}
Koslovsky, M.~D. (2023).
\newblock A bayesian zero-inflated dirichlet-multinomial regression model for multivariate compositional count data.
\newblock {\em Biometrics}, 79(4):3239--3251.

\bibitem[Lambert, 1992]{Lambert}
Lambert, D. (1992).
\newblock Zero-inflated poisson regression, with an application to defects in manufacturing.
\newblock {\em Technometrics}, 34(1):1--14.

\bibitem[Louzada et~al., 2018]{Louzada2017}
Louzada, F., Moreira, F.~F., and de~Oliveira, M.~R. (2018).
\newblock A zero-inflated non default rate regression model for credit scoring data.
\newblock {\em Communications in Statistics - Theory and Methods}, 47(12):3002--3021.

\bibitem[Martin et~al., 2005]{Martins}
Martin, T.~G., Wintle, B.~A., Rhodes, J.~R., Kuhnert, P.~M., Field, S.~A., Low-Choy, S.~J., Tyre, A.~J., and Possingham, H.~P. (2005).
\newblock Zero tolerance ecology: improving ecological inference by modelling the source of zero observations.
\newblock {\em Ecology Letters}, 8(11):1235--1246.

\bibitem[Menezes et~al., 2025]{menezes2025}
Menezes, A. F.~B., Parnell, A.~C., and Murphy, K. (2025).
\newblock Finite mixture representations of zero-\&-$n$-inflated distributions for count-compositional data.

\bibitem[Mullahy, 1986]{Mullahy1986}
Mullahy, J. (1986).
\newblock Specification and testing of some modified count data models.
\newblock {\em Journal of Econometrics}, 33(3):341--365.

\bibitem[Ospina and Ferrari, 2012]{Ospina2012}
Ospina, R. and Ferrari, S. L.~P. (2012).
\newblock A general class of zero-or-one inflated beta regression models.
\newblock {\em Computational Statistics \& Data Analysis}, 56(6):1609--1623.

\bibitem[Parnell et~al., 2014]{Parnell}
Parnell, A.~C., Sweeney, J., Doan, T.~K., Salter-Townshend, M., Allen, J. R.~M., Huntley, B., and Haslett, J. (2014).
\newblock Bayesian inference for palaeoclimate with time uncertainty and stochastic volatility.
\newblock {\em Journal of the Royal Statistical Society Series C: Applied Statistics}, 64(1):115--138.

\bibitem[Pereira et~al., 2013]{Pereira2013}
Pereira, G.~H., Botter, D.~A., and Sandoval, M.~C. (2013).
\newblock A regression model for special proportions.
\newblock {\em Statistical Modelling}, 13(2):125--151.

\bibitem[Royle and Dorazio, 2008]{Royle}
Royle, J.~A. and Dorazio, R.~M. (2008).
\newblock {\em Hierarchical Modeling and Inference in Ecology: The Analysis of Data from Populations, Metapopulations and Communities}.
\newblock Elesevier Academic Press.

\bibitem[Royle and Link, 2006]{Royle2006}
Royle, J.~A. and Link, W.~A. (2006).
\newblock Generalized site occupancy models allowing for false positive and false negative errors.
\newblock {\em Ecology}, 87(4):835--841.

\bibitem[Salter-Townshend and Haslett, 2012]{MST}
Salter-Townshend, M. and Haslett, J. (2012).
\newblock Fast inversion of a flexible regression model for multivariate pollen counts data.
\newblock {\em Environmetrics}, 23:595--605.

\bibitem[{Stan Development Team}, 2025]{rstan}
{Stan Development Team} (2025).
\newblock {RStan}: the {R} interface to {Stan}.
\newblock R package version 2.32.7.

\bibitem[Tian et~al., 2015]{Tian2015}
Tian, G.-L., Ma, H., Zhou, Y., and Deng, D. (2015).
\newblock Generalized endpoint-inflated binomial model.
\newblock {\em Computational Statistics \& Data Analysis}, 89:97--114.

\bibitem[Tomarchio and Punzo, 2019]{Tomarchio2019}
Tomarchio, S.~D. and Punzo, A. (2019).
\newblock Modelling the loss given default distribution via a family of zero-and-one inflated mixture models.
\newblock {\em Journal of the Royal Statistical Society Series A: Statistics in Society}, 182(4):1247--1266.

\bibitem[Wright et~al., 2017]{Wright2017}
Wright, W.~J., Irvine, K.~M., Warren, J.~M., and Barnett, J.~K. (2017).
\newblock Statistical design and analysis for plant cover studies with multiple sources of observation errors.
\newblock {\em Methods in Ecology and Evolution}, 8(12):1832--1841.

\end{thebibliography}

\end{document}